\documentclass[aps,prd,twocolumn,showpacs,superscriptaddress]{revtex4-1}  
\usepackage{graphicx}  
\usepackage{dcolumn}   
\usepackage{bm}        
\usepackage{amssymb}   
\usepackage{amsmath}

\usepackage{subfigure}
\usepackage{float}
\usepackage{color}
\usepackage[toc,page]{appendix}
\usepackage{graphicx}
\usepackage{booktabs}
\usepackage{hyperref}
\usepackage{multirow}
\usepackage{enumitem}
\usepackage{soul}

\usepackage[none]{hyphenat}

\newcommand{\yield}{(5.91\pm0.09)\times10^{-43}~\rm{cm}^2/\rm{fission}}
\newcommand{\RHuber}{0.952\pm0.014\pm0.023}
\newcommand{\RILL}{1.001\pm0.015\pm0.027}

\begin{document}



\title{
Improved Measurement of the Reactor Antineutrino Flux at Daya Bay
}

\newcommand{\ECUST}{\affiliation{Institute of Modern Physics, East China University of Science and Technology, Shanghai}}
\newcommand{\IHEP}{\affiliation{Institute~of~High~Energy~Physics, Beijing}}
\newcommand{\Wisconsin}{\affiliation{University~of~Wisconsin, Madison, Wisconsin 53706}}
\newcommand{\Yale}{\affiliation{Wright~Laboratory and Department~of~Physics, Yale~University, New~Haven, Connecticut 06520}} 
\newcommand{\BNL}{\affiliation{Brookhaven~National~Laboratory, Upton, New York 11973}}
\newcommand{\NTU}{\affiliation{Department of Physics, National~Taiwan~University, Taipei}}
\newcommand{\NUU}{\affiliation{National~United~University, Miao-Li}}
\newcommand{\Dubna}{\affiliation{Joint~Institute~for~Nuclear~Research, Dubna, Moscow~Region}}
\newcommand{\CalTech}{\affiliation{California~Institute~of~Technology, Pasadena, California 91125}}
\newcommand{\CUHK}{\affiliation{Chinese~University~of~Hong~Kong, Hong~Kong}}
\newcommand{\NCTU}{\affiliation{Institute~of~Physics, National~Chiao-Tung~University, Hsinchu}}
\newcommand{\NJU}{\affiliation{Nanjing~University, Nanjing}}
\newcommand{\TsingHua}{\affiliation{Department~of~Engineering~Physics, Tsinghua~University, Beijing}}
\newcommand{\SZU}{\affiliation{Shenzhen~University, Shenzhen}}
\newcommand{\NCEPU}{\affiliation{North~China~Electric~Power~University, Beijing}}
\newcommand{\Siena}{\affiliation{Siena~College, Loudonville, New York  12211}}
\newcommand{\IIT}{\affiliation{Department of Physics, Illinois~Institute~of~Technology, Chicago, Illinois  60616}}
\newcommand{\LBNL}{\affiliation{Lawrence~Berkeley~National~Laboratory, Berkeley, California 94720}}
\newcommand{\UIUC}{\affiliation{Department of Physics, University~of~Illinois~at~Urbana-Champaign, Urbana, Illinois 61801}}
\newcommand{\SJTU}{\affiliation{Department of Physics and Astronomy, Shanghai Jiao Tong University, Shanghai Laboratory for Particle Physics and Cosmology, Shanghai}}
\newcommand{\BNU}{\affiliation{Beijing~Normal~University, Beijing}}
\newcommand{\WM}{\affiliation{College~of~William~and~Mary, Williamsburg, Virginia  23187}}
\newcommand{\Princeton}{\affiliation{Joseph Henry Laboratories, Princeton~University, Princeton, New~Jersey 08544}}
\newcommand{\VirginiaTech}{\affiliation{Center for Neutrino Physics, Virginia~Tech, Blacksburg, Virginia  24061}}
\newcommand{\CIAE}{\affiliation{China~Institute~of~Atomic~Energy, Beijing}}
\newcommand{\SDU}{\affiliation{Shandong~University, Jinan}}
\newcommand{\NanKai}{\affiliation{School of Physics, Nankai~University, Tianjin}}
\newcommand{\UC}{\affiliation{Department of Physics, University~of~Cincinnati, Cincinnati, Ohio 45221}}
\newcommand{\DGUT}{\affiliation{Dongguan~University~of~Technology, Dongguan}}
\newcommand{\XJTU}{\affiliation{Department of Nuclear Science and Technology, School of Energy and Power Engineering, Xi'an Jiaotong University, Xi'an}}
\newcommand{\UCB}{\affiliation{Department of Physics, University~of~California, Berkeley, California  94720}}
\newcommand{\HKU}{\affiliation{Department of Physics, The~University~of~Hong~Kong, Pokfulam, Hong~Kong}}
\newcommand{\UH}{\affiliation{Department of Physics, University~of~Houston, Houston, Texas  77204}}
\newcommand{\Charles}{\affiliation{Charles~University, Faculty~of~Mathematics~and~Physics, Prague}} 
\newcommand{\USTC}{\affiliation{University~of~Science~and~Technology~of~China, Hefei}}
\newcommand{\TempleUniversity}{\affiliation{Department~of~Physics, College~of~Science~and~Technology, Temple~University, Philadelphia, Pennsylvania  19122}}
\newcommand{\CUC}{\affiliation{Instituto de F\'isica, Pontificia Universidad Cat\'olica de Chile, Santiago}} 
\newcommand{\CGNPG}{\affiliation{China General Nuclear Power Group, Shenzhen}}
\newcommand{\NUDT}{\affiliation{College of Electronic Science and Engineering, National University of Defense Technology, Changsha}} 
\newcommand{\IowaState}{\affiliation{Iowa~State~University, Ames, Iowa  50011}}
\newcommand{\ZSU}{\affiliation{Sun Yat-Sen (Zhongshan) University, Guangzhou}}
\newcommand{\CQU}{\affiliation{Chongqing University, Chongqing}} 
\newcommand{\BCC}{\altaffiliation[Now at ]{Department of Chemistry and Chemical Technology, Bronx Community College, Bronx, New York  10453}} 
\author{D.~Adey}\IHEP
\author{F.~P.~An}\ECUST
\author{A.~B.~Balantekin}\Wisconsin
\author{H.~R.~Band}\Yale
\author{M.~Bishai}\BNL
\author{S.~Blyth}\NTU\NUU
\author{D.~Cao}\NJU
\author{G.~F.~Cao}\IHEP
\author{J.~Cao}\IHEP
\author{Y.~L.~Chan}\CUHK
% ...

\date{\today}

\begin{abstract}
  This work reports a precise measurement of the reactor
  antineutrino flux using 2.2 million inverse beta decay (IBD) events
  collected with the Daya Bay near detectors in 1230 days.  The
  dominant uncertainty on the neutron detection efficiency is
  reduced by 56\% with respect to the previous measurement through a comprehensive neutron
  calibration and detailed data and simulation analysis.
  The new average IBD yield is determined to be $\yield$ with total
  uncertainty improved by 29\%.  The corresponding mean fission
  fractions from the four main fission isotopes $^{235}$U, $^{238}$U,
  $^{239}$Pu, and $^{241}$Pu are 0.564, 0.076, 0.304, and 0.056,
  respectively.  The ratio of measured to predicted antineutrino yield
  is found to be $\RHuber$~($\RILL$) for
  the Huber-Mueller (ILL-Vogel) model, where the first and second
  uncertainty are experimental and theoretical model uncertainty,
  respectively. This measurement confirms the discrepancy
  between the world average of reactor antineutrino flux and the Huber-Mueller model.
\end{abstract}

\pacs{}
\maketitle

\section{Introduction}
Nuclear reactors are an intense man-made source of electron antineutrinos
and were used for the first observation of the neutrino~\cite{Cowan:1992xc}.
Electron antineutrinos can be detected
through inverse beta decay (IBD) on target protons, where a prompt
positron and a delayed neutron capture signals are measured in time
coincidence.  Since the early 2000s, the energy and baseline (the distance between source and detector) dependent
neutrino disappearance at nuclear reactors~\cite{Eguchi:2002dm,
  An:2012eh, Ahn:2012nd}
has provided strong evidence of neutrino oscillation~\cite{Pontecorvo:1957cp, Pontecorvo:1967fh,
  Maki:1962mu}.  However, a recent re-evaluation of the theoretical
prediction (referred to as Huber-Mueller model~\cite{Huber, Mueller})
of the reactor neutrino flux resulted in a $\sim$6\% deficit in
measured flux from short-baseline experiments~\cite{Mention:2011rk}
and the previous ILL-Vogel model~\cite{ILL1, ILL2, ILL3, Vogel}.  The
difference between the data and Huber-Mueller prediction, i.e.~the
so-called ``reactor antineutrino anomaly'' (RAA), could be interpreted
as active-to-sterile neutrino oscillation with a mass-squared
splitting ($\Delta m^2$) around 1~eV$^2$. It is also shown in
Refs.~\cite{Kopp:2011qd, Kopp:2013vaa, Giunti:2013aea} that the
allowed parameter space is compatible with earlier anomalies from
LSND~\cite{Athanassopoulos:1997pv, Aguilar:2001ty},
MiniBooNE~\cite{AguilarArevalo:2010wv}, GALLEX~\cite{Hampel:1997fc},
and SAGE~\cite{Abdurashitov:2009tn}.  On the other hand, a number of
authors~\cite{Giunti, Dentler, Giunti2, Gebre} have argued that the
RAA may be due to the theoretical uncertainties in the flux
calculations. Recent antineutrino flux evolution results from Daya Bay are
in tension with the sterile-neutrino-only explanation of
RAA~\cite{An:2017osx}.

The uncertainty of the reactor antineutrino flux in our previous measurement~\cite{An:2017osx}
is dominated by the uncertainty of neutron detection efficiency.
The neutron detection efficiency was determined to be $\varepsilon_{n}=(81.83\pm1.38)\%$~\cite{An:2015nua, An:2016srz},
and the ratio with respect to the total uncertainty is $\sigma_{\varepsilon_{n}}^2/\sigma_{total}^2=65\%$.
To further elucidate the RAA situation, this work presents an updated
flux measurement from Daya Bay using the same 1230-day data set, but
with a more precise determination of the neutron detection
efficiency. Key improvements include an elaborated neutron
calibration campaign covering a wide range of neutron energy and
positions, an improved simulation with different physics models, and
a data-driven correction to the neutron efficiency.

This paper is organized as follows.  In Sec.~\ref{sec:method}, we
explain the general method to measure reactor neutrino yield, and
highlight our approach here to improve its estimate.
Sec.~\ref{sec:analysis} discusses the neutron calibration campaign and
the analysis of calibration and simulation data.  In
Sec.~\ref{sec:result} we present an improved reactor antineutrino flux measurement
and a comparison with the world data and theoretical models.

\section{Method}
\label{sec:method}
\subsection{Overview of procedure}
\begin{figure}[]
    \centering
    \includegraphics[width=0.4\textwidth]{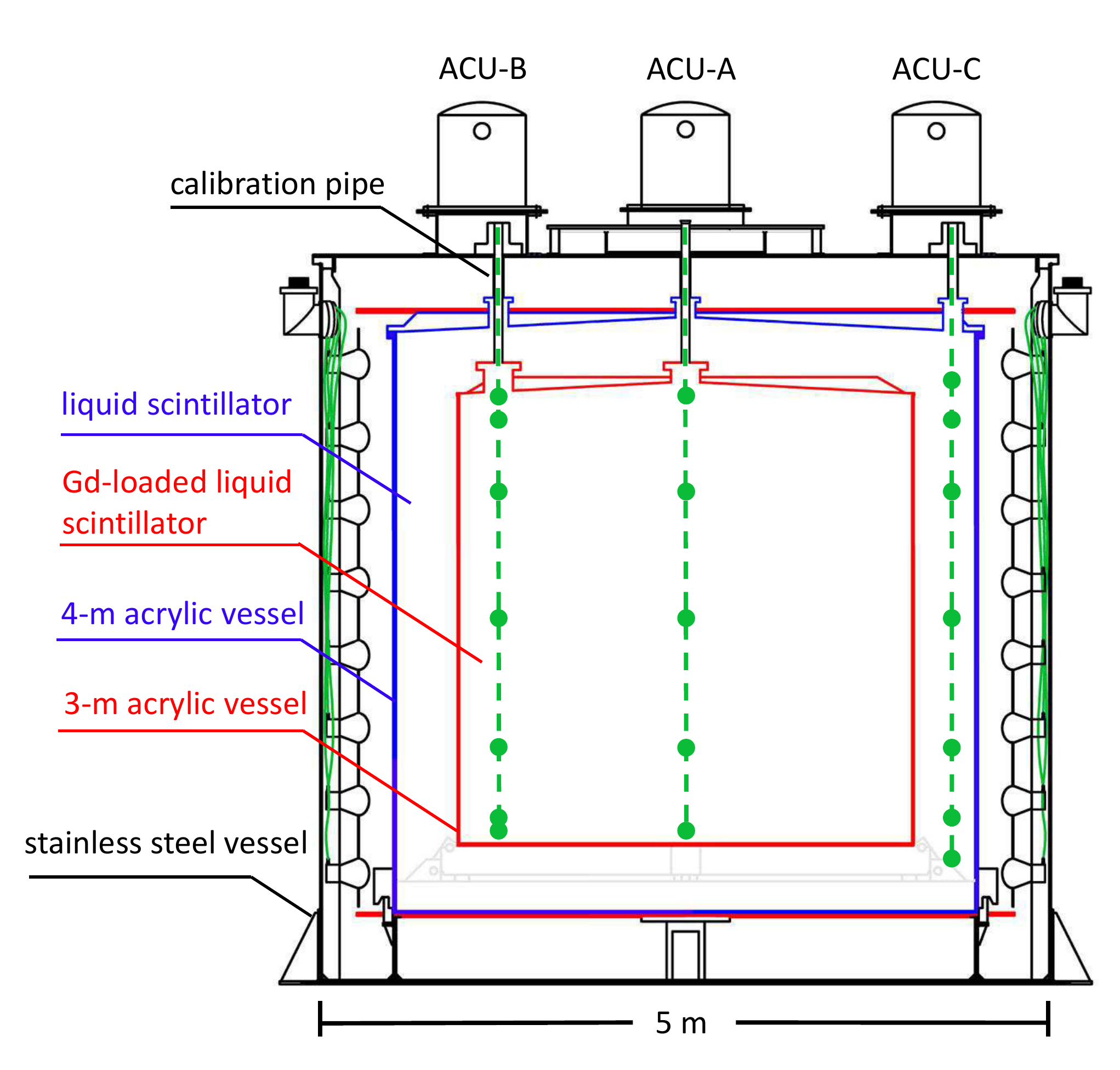}
    \caption{Schematic diagram of the Daya Bay antineutrino detector.
    On the top, three automated calibration unit (ACU-A, B, and C), are installed.
    The three related vertical calibration axes are shown as the dashed lines, with all the
    locations of the calibration points highlighted.}
    \label{fig:det}
\end{figure}

The Daya Bay experiment has four near and four far identically designed
antineutrino detectors (ADs), located at different baselines
(360~m--1900~m)~\cite{An:2015qga} measuring the electron antineutrino flux
from six reactor cores.
The structure of the detector is shown in Figure~\ref{fig:det}.
Each AD consists of a cylindrical target volume with 20 tons of
0.1\% gadolinium loaded liquid scintillator (GdLS, 3.1 m in diameter and 3.1 m in
height), surrounded by a layer of 42-cm thick liquid
scintillator (LS) to enclose the gammas or electrons escaped from the central GdLS
region.  The GdLS and LS are separated by a 1-cm thick acrylic vessel.
An energy deposit in the GdLS and LS regions is detected by photomultiplier tubes (PMTs).
The origin of the coordinate system is set at the
geometrical center of the GdLS cylinder, with the $z$-axis pointing up.
IBD neutrons are detected by delayed capture either on hydrogen emitting one 2.2 MeV
gamma or on gadolinium emitting several gammas with total energy of about 8 MeV.
The kinetic energy of the IBD neutrons is less than 50 keV.
The average capture time in the GdLS region is about 28.5~$\mu$s and
216~$\mu$s in the LS~\cite{nHlong}.

The reactor antineutrino IBD candidates are selected with the same criteria
as in Ref.~\cite{An:2016ses}, which are also described here. 1) Removal of events caused by PMT light emission.
2) The time between the prompt and delayed signal is in the range of [1, 200]~$\mu$s.
3) Prompt signal must have a reconstructed energy, E, between 0.7 and 12 MeV.
4) Delayed signal must have E between 6 and 12 MeV to select
gadolinium captures.  5) Muon anti-coincidence.  6) Multiplicity cut
to remove events with E $>$ 0.7 MeV in the
interval 200 $\mu$s before the prompt signal, 200 $\mu$s after
the delayed signal, or between the prompt and delayed signals.
The dominant backgrounds are accidental coincident events and
cosmic-ray muon induced $^9$Li/$^8$He, which are less than 2\% of the
signal IBD rate for the four near ADs.
After statistical subtraction of background, the total number of IBD
signals, $N_{\text{IBD}}$, is $2.201\times10^6$ for the four near detectors.

To compare to the theoretical predictions, the reactor antineutrino yield
$\sigma_f$, defined as the number of antineutrinos times IBD cross-section per fission, can
be calculated by solving the following equation:
\begin{equation}
\label{eq:yieldcal}
N_{\text{IBD}}(1-c^{\rm{SNF}}) = \sigma_f \sum_{d=1}^4\sum_{r=1}^6\frac{N_d^P \varepsilon_{\text{IBD}}P_{sur}^{rd} N_r^f}{4\pi L_{rd}^2},
\end{equation}
where the index $d$ is for four near detectors, index $r$ is for the
six reactor cores, $N_d^P$ is the number of target protons of detector
$d$, $\varepsilon_{\text{IBD}}$ is the IBD detection efficiency,
$P_{sur}^{rd}$ is the mean neutrino survival probability from the
reactor $r$ to detector $d$, $N_r^f$ is the predicted number of
fissions of the $r$th reactor core,
$L_{rd}$ is the distance from reactor $r$ to detector $d$,
and $c^{\text{SNF}}$ is a correction term for spent nuclear fuel.
$P_{sur}^{rd}$ is calculated by integrating the
cross-section-weighted oscillation survival probability over the
$\bar{\nu}_e$ energy spectrum, using $\sin^22\theta_{13}$ and $|\Delta
m_{ee}^2|$ determined from the same
data~\cite{An:2016ses}.
The average oscillation correction for
near detectors is 1.5\%.

The IBD detection efficiency is divided into two factors:
\begin{equation}
\label{eq:eff_IBD}
\varepsilon_{\rm{IBD}} = \varepsilon_n \times \varepsilon_{\rm{other}},
\end{equation}
where $\varepsilon_n$ is the neutron selection efficiency due to the [6, 12] MeV cut and
$\varepsilon_{\rm{other}}$ is for the PMT light emission, prompt energy,
and coincident time cuts.

The predicted number of fissions of the $r$th reactor core is
\begin{equation}
\label{eq:Nfission}
N_r^f = \int\frac{W_r}{\sum_{iso=1}^4 f_r^{iso} E^{iso}}{\rm{d}}t,
\end{equation}
where $W_r$ is the thermal power of the $r$th core,
$E^{iso}$ is the mean energy released per fission for each isotope,
and $f_r^{iso}$ is the average fission fraction of the $r$th core for each isotope,
and the ratio is integrated over the live time of the detectors.
The original thermal power and fission fuel composition data are provided by the power plant.
$c^{\text{SNF}}$ was estimated to be $(0.3\pm0.3)\%$ previously~\cite{An:2016srz}.

Different components of relative uncertainties for the antineutrino
yield measurement from previous work~\cite{An:2015nua, An:2016srz}, including
$\varepsilon_n$ and $\varepsilon_{\rm{other}}$, are summarized in
Table~\ref{tab:Effi}. Clearly $\varepsilon_n$ dominates the uncertainty,
and is the target of improvement in this paper.

\begin{table}[]
\centering
\caption{Summary of relative uncertainties for the flux measurements and the measured flux to model prediction ratio measurements
  in our previous study~\cite{An:2016srz} and this work.
  Central values of the detector efficiencies, $\varepsilon_n$ and $\varepsilon_{\rm{other}}$, are also listed.}
\label{tab:Effi}
\begin{tabular}{lcccc}
  \hline\hline
                    & \multicolumn{2}{c}{Previous} & \multicolumn{2}{c}{This work} \\
  source            & value   & rel. err.   & value     & rel. err.     \\
  \hline
  statistic         &-        &  0.1\%      & -      &  0.1\%           \\
  oscillation       &-        &  0.1\%      & -      &  0.1\%           \\
  target proton     &-        &  0.92\%     & -      &  0.92\%          \\
  reactor  \\
    \hspace*{0.5cm} power            &-        &  0.5\% &-        &  0.5\% \\
    \hspace*{0.5cm} energy/fission   &-        &  0.2\% &-        &  0.2\% \\
    \hspace*{0.5cm} IBD cross section&-        &  0.12\%&-        &  0.12\%\\
    \hspace*{0.5cm} fission fraction &-        &  0.6\% &-        &  0.6\% \\
    \hspace*{0.5cm} spent fuel       &-        &  0.3\% &-        &  0.3\% \\
    \hspace*{0.5cm} non-equilibrium  &-        &  0.2\% &-        &  0.2\% \\
  $\varepsilon_{\rm{IBD}}$ \\
    \hspace*{0.5cm} $\varepsilon_n$ & 81.83\% &  1.69\%     & 81.48\%   &  0.74\%            \\
    \hspace*{0.5cm} $\varepsilon_{\rm{other}}$  & 98.49\% &  0.16\%     & 98.49\% &  0.16\%  \\
  \hline
  total             & -       &  2.1\%      & -         &  1.5\%        \\
  \hline\hline
\end{tabular}
\end{table}

\subsection{Principle of improvement}
The neutron detection efficiency, $\varepsilon_n$, is composed of three
individual factors.
\begin{itemize}
\item  The Gd capture fraction is the fraction of neutrons produced by IBD in the
  GdLS target that are captured on Gd.
  The capture fraction is lower at the edge of GdLS volume because neutrons may
  drift into un-doped LS volume (spill-out effect).
\item The nGd gamma detection efficiency is the fraction of neutron Gd capture signals with detected gamma energy above 6 MeV.
\item Spill-in: efficiency increase due to IBD events produced in the LS and acrylic but
  with neutron capture on Gd.
\end{itemize}
We note that the estimation of Gd capture fraction and spill-in effects are strongly
correlated since they are both driven by the modeling of neutron
propagation including neutron scattering in materials and the subsequent
nuclear capture.
The estimation of the nGd gamma detection efficiency depends on the modeling
of gamma emission including the multiplicity and energy spectrum of the
emitted gammas.

In the previous study~\cite{An:2016srz}, we attempted to use different
neutron calibration data to estimate these individual effects.
The main difficulty was that no data can cleanly separate their uncertainties.
In this paper, instead, we evaluate $\varepsilon_n$ and its
uncertainty directly using a new neutron calibration data set and a
data-simulation comparison. This approach is data driven and allows a
significant reduction of the uncertainty.


\section{Improved Detection Efficiency Estimation}
\label{sec:analysis}
\subsection{Neutron calibration campaign}
An extensive neutron calibration campaign was carried out in Daya Bay
at the end of 2016.  Two types of custom sources were fabricated,
$^{241}$Am-$^{13}$C (AmC, neutron rate $\sim$ 100
Hz)~\cite{Liu:2015cra} and $^{241}$Am-$^9$Be (AmBe, neutron rate
$\sim$ 30 Hz). They produce neutrons through
$^{13}$C($\alpha$, n)$^{16}$O or $^{9}$Be($\alpha$, n)$^{12}$C
reactions with the final nucleus either in the ground state
(GS) or excited state (ES). The kinetic energy of the neutrons
from AmC (AmBe) in the GS and ES are [3, 7] MeV ([6, 10] MeV) and $<$1 MeV ([2, 6] MeV), respectively.
Calibration events are formed from the prompt energy of the proton recoil
and deexcitation gammas, if $^{16}$O$^*$ or $^{12}$C$^*$ is created,
and the delayed neutron capture.
The high neutron rate and delayed-time-coincidence present a high
signal-to-background ratio for the calibration study.

The sources, sealed in a small stainless steel cylinder (8 mm in both
diameter and height), were enclosed in a highly reflective
PTFE (Polytetrafluoroethylene) shell.
These sources were deployed vertically into a near-site AD using the automated calibration units
(ACUs)~\cite{Liu:2013ava} along the central axis (ACU-A), an edge
axis of GdLS at a radius of 1.35~m (ACU-B), and through a middle axis of
the LS layer at a radius of 1.77~m (ACU-C). During deployment, the
absolute precision of the source $z$ location is 7~mm~\cite{Liu:2013ava}.
All calibration positions are illustrated in
Figure~\ref{fig:det}.  In total, data in 59 different
source (final nucleus states) and location points (SLPs), were collected.

\begin{figure}[h!]
    \centering
    \includegraphics[width=0.4\textwidth]{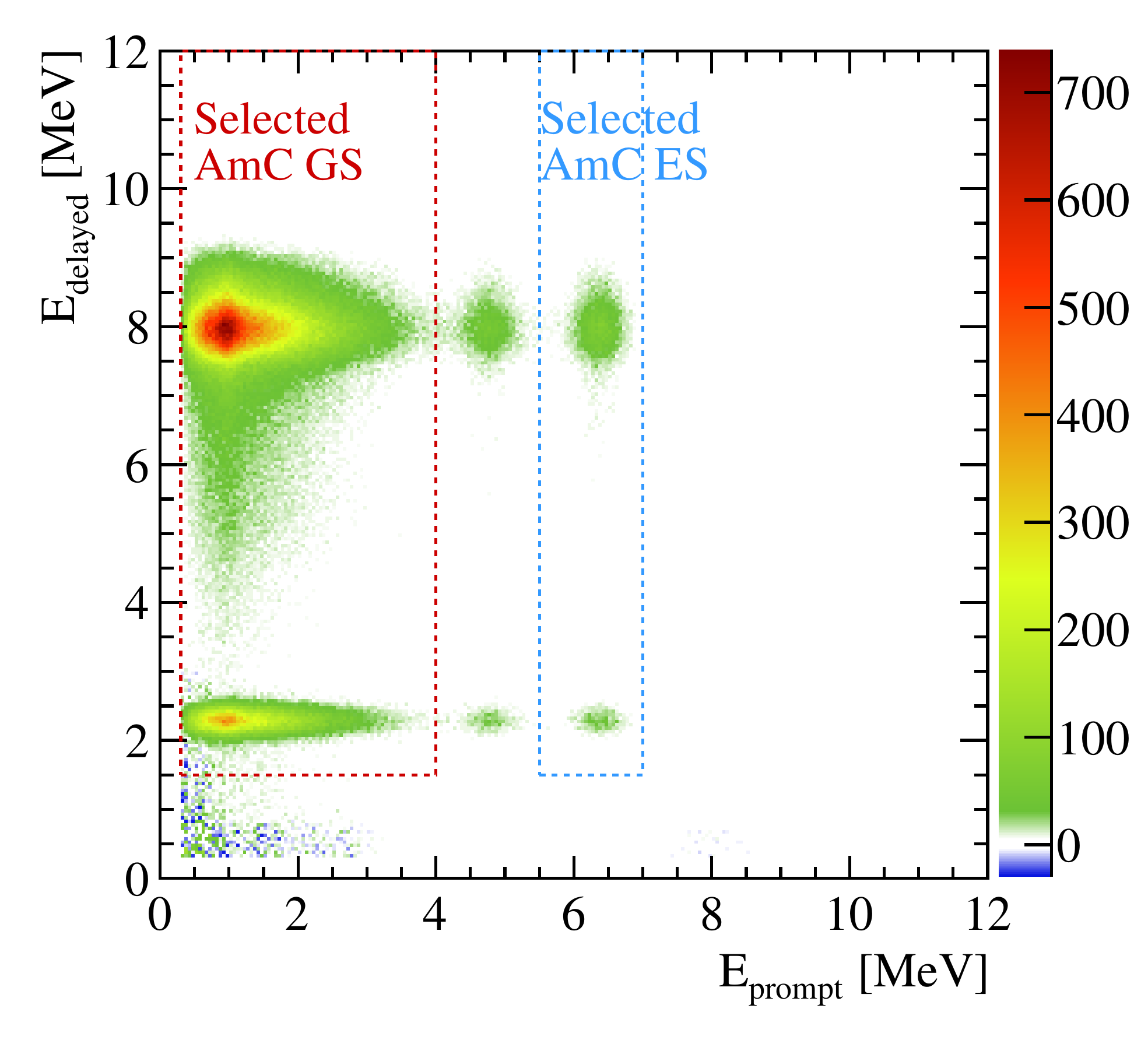}
    \caption{The prompt vs.~delayed energy distribution of one AmC sample (ACU-B $z$=0 m).
    The selected ground state (GS) and excited state (ES) are indicated.
    The clusters in the prompt energy spectrum between [4, 5.5] MeV are caused by $^{12}$C(n,n$\gamma$)$^{12}$C, and are not used.
    Negative bin content is due to background subtraction.}
    \label{fig:AmC}
    \includegraphics[width=0.4\textwidth]{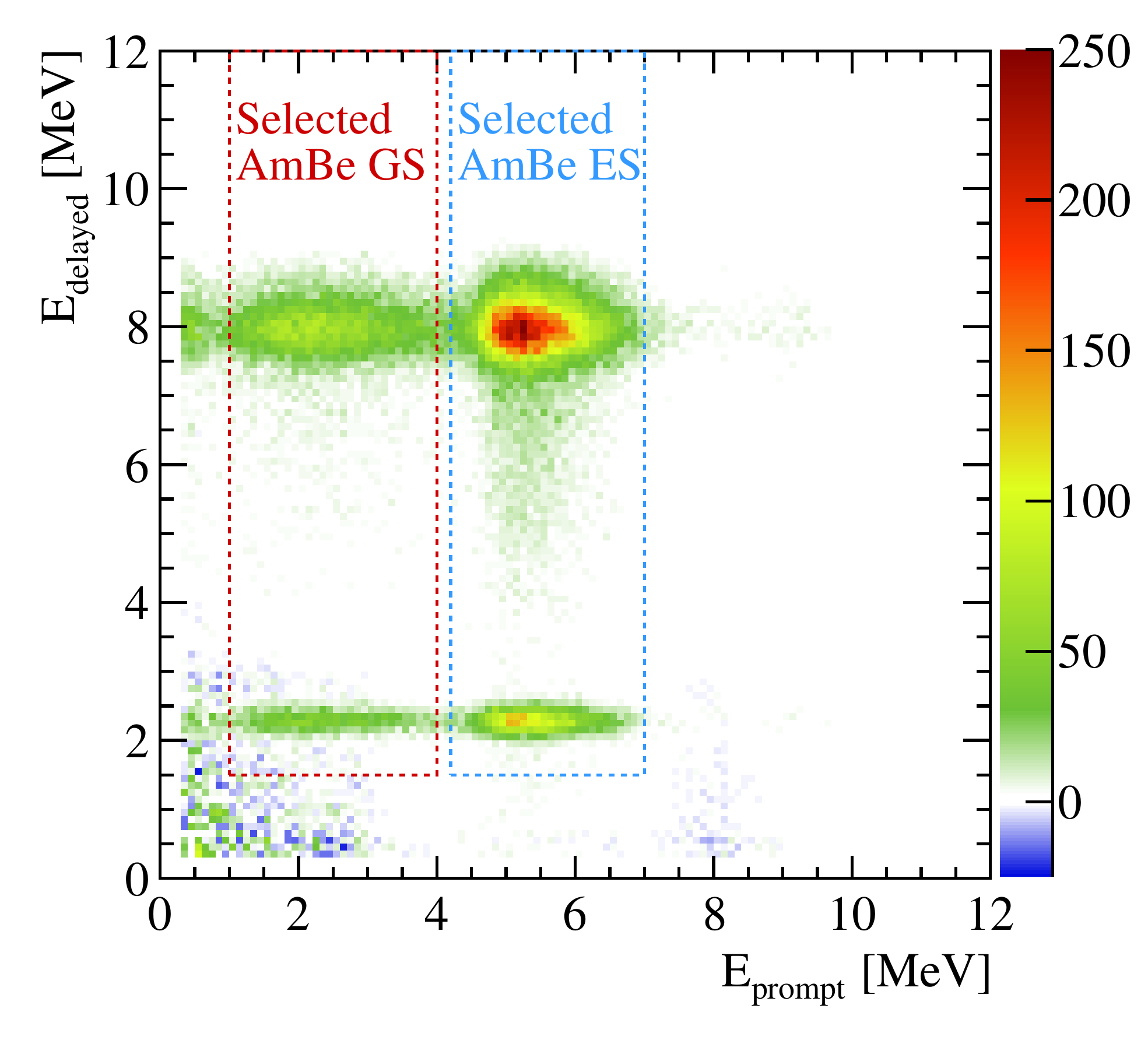}
    \caption{The prompt vs.~delayed energy distribution of one AmBe sample (ACU-B $z$=0 m).
    The selected ground state (GS) and excited state (ES) are indicated.
    Negative bin content is due to background subtraction.}
    \label{fig:AmBe}
\end{figure}

Delayed coincidence events for the calibration sample are selected with a
time requirement of $1~\mu\rm{s} <\Delta t < 1200~\mu\rm{s}$ for all
events with E between $[0.3, 12]~\rm{MeV}$.  The $1200~\mu\rm{s}$ selection cut is set
to efficiently include neutron captures in the LS and acrylic region.
Two example distributions of the prompt-delay energies of AmC (ACU-B
$z$=0 m) and AmBe (ACU-B $z$=0 m) samples are shown in Figure~\ref{fig:AmC} and
Figure~\ref{fig:AmBe}, respectively.  The data in different channels are
selected using the following prompt energy cut:
$[0.3, 4]~\rm{MeV}$ for AmC GS, $[5.5, 7]~\rm{MeV}$ for AmC ES,
$[1, 4]~\rm{MeV}$ for AmBe GS, and $[4.2, 7]~\rm{MeV}$ for AmBe ES.
The accidental background contributes 0.1--20\% of the neutron candidates depending on the SLP, and is estimated
by randomly paired single events~\cite{nHlong}.
The reactor antineutrino and cosmogenic backgrounds are estimated by applying the
same selection cuts on the data acquired immediately before
and after the calibration campaign.
They contribute $<$0.1\% to the neutron source signals. All of these backgrounds are statistically subtracted.

\subsection{Neutron and gamma modeling in simulation}
The neutron calibration data were compared to the model
predictions obtained using the Geant4-based~\cite{Agostinelli:2002hh} (v4.9.2)
Daya Bay Monte Carlo (MC) simulation framework NuWa~\cite{An:2012eh, An:2015qga}
with improvements to
the calibration pipe geometry, detector energy response, and neutron
transport modelling~\footnote{
An error in neutron propagation was corrected. Details can be found in \url{https://bugzilla-geant4.kek.jp/show_bug.cgi?id=1856}. We replaced the G4UHadronElasticProcess with G4HadronElasticProcess to correct the error, and the IBD selection efficiency increased by 0.3\%.}.
These modifications improve the agreement between the neutron
calibration data~\cite{An:2016ses} and simulation.

\begin{figure}[]
    \centering
    \includegraphics[width=0.45\textwidth]{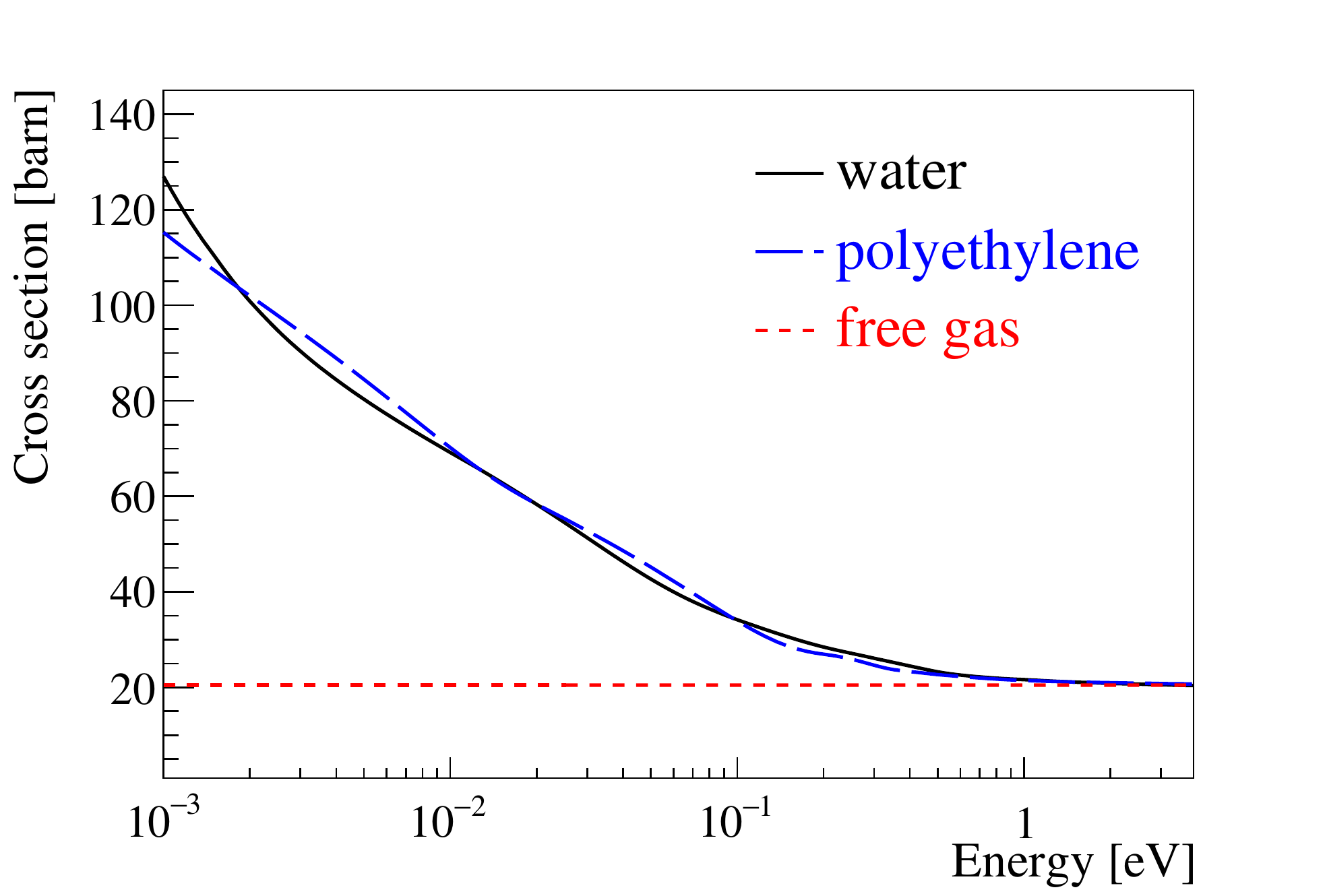}
    \caption{Total scattering cross-section as a function of neutron kinetic energy for three models.
    The data are extracted from Geant4 simulation.}
    \label{fig:neutron}
\end{figure}

Neutrons lose energy through various scattering processes before capture on a nucleus.
There are no scattering models in Geant4 for the Daya Bay scintillator
(average hydrogen-to-carbon ratio CH$_{1.61\sim1.64}$) or acrylic (C$_5$O$_2$H$_8$).
Above 4 eV, a generic Geant4 model, ``G4NeutronHPElastic'' can be selected for neutron simulation.
Below 4 eV, three possible options from the Geant4 data libraries are an
elastic scattering model without molecular bonds (``free gas''), a water
model (H$_2$O), and a polyethylene model (CH$_2$, ``poly'')~\cite{Agostinelli:2002hh}.
The latter two models are built based on
the ENDF database~\cite{Chadwick:2011xwu} and are quite different from
the free gas model.  The total scattering cross-section as a function of
energy for the three models is shown in Figure~\ref{fig:neutron}.  To
approximate the Daya Bay (scintillator, acrylic) material pair, five
combinations (Table~\ref{tab:models}) of models were studied, including a) (water, free gas) b)
(water, poly), c) (poly, poly), d) (poly, free gas), and e) (free gas,
free gas).

\begin{figure}[]
    \centering
    \subfigure[$^{155}$Gd]
    {
      \label{fig:gamma155}
      \includegraphics[width=0.47\textwidth]{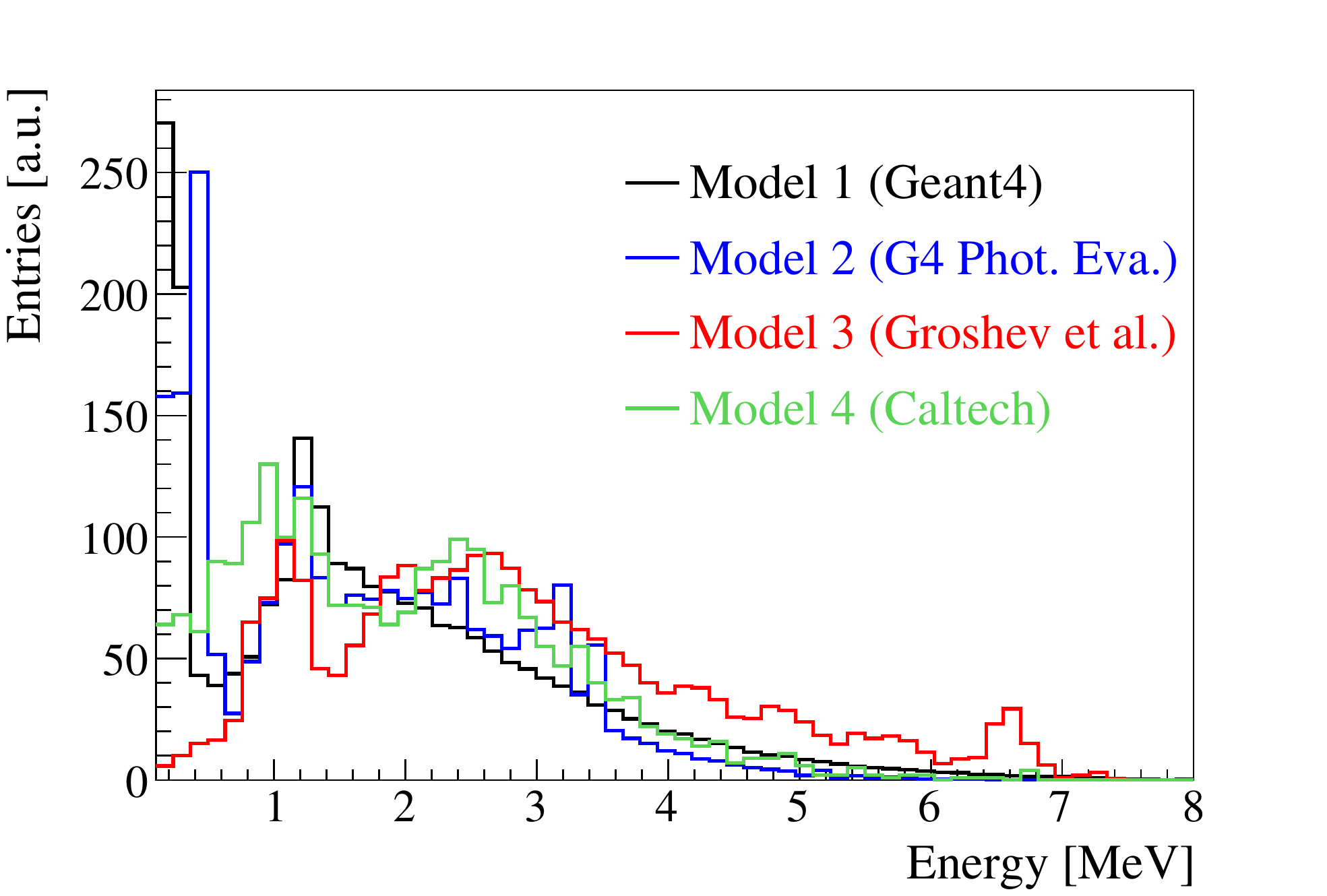}
    }
    \subfigure[$^{157}$Gd]
    {
      \label{fig:gamma157}
      \includegraphics[width=0.47\textwidth]{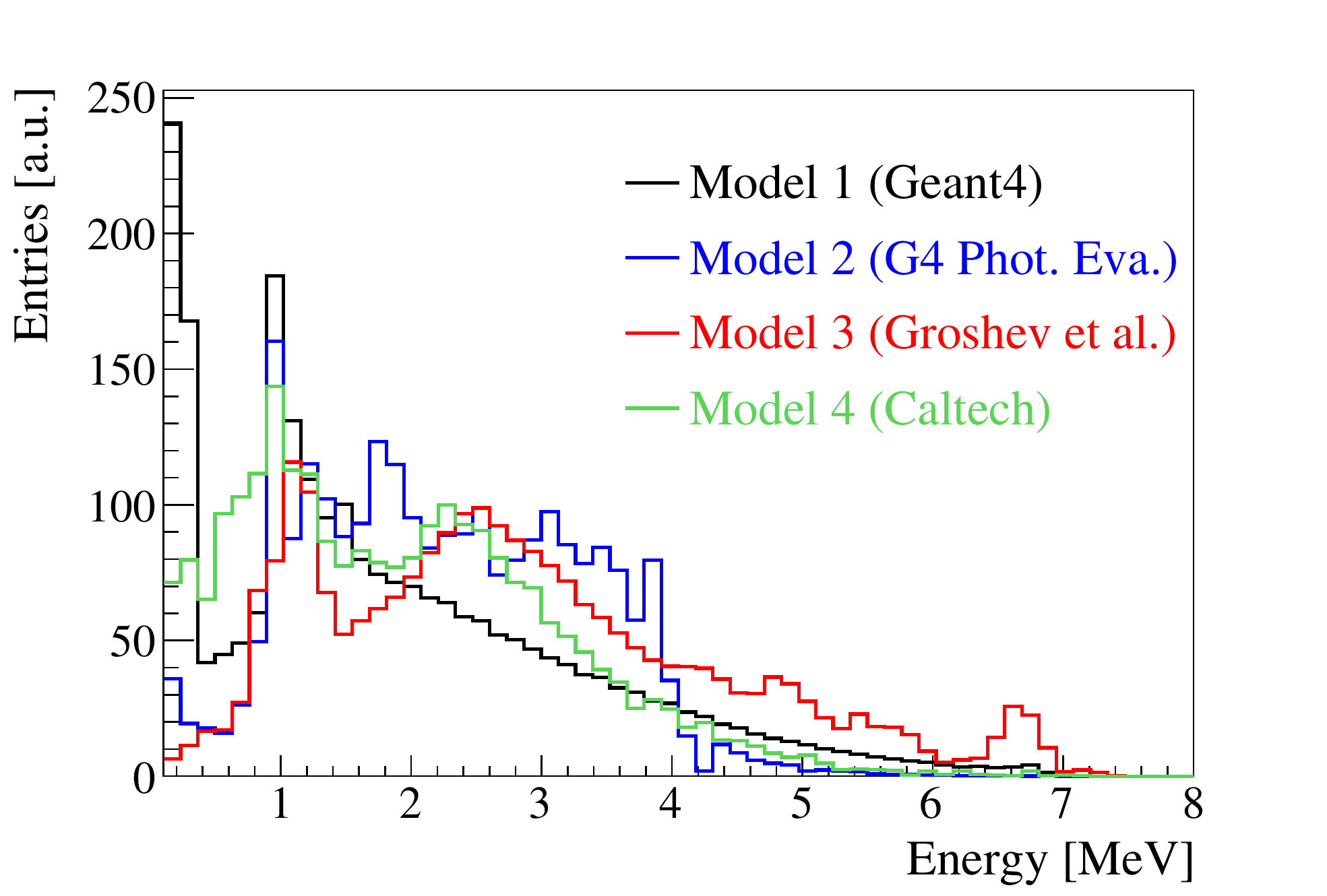}
    }
    \caption{Energy distribution of the deexcitation gammas of $^{155}$Gd (a) and $^{157}$Gd (b) for the four models as indicated in the legend.}
\end{figure}

For the neutron capture gamma energy and multiplicity distributions,
four different models (Table~\ref{tab:models}) were selected, including 1) a native
Geant4 model, 2) a Geant4 model with the photon evaporation process,
3) a model based on the Nuclear Data Sheets by L.~Groshev~{\it{et al.}}~\cite{groshev1968compendium},
and 4) a model based on the measured single gamma distribution of nGd
capture at Caltech~\cite{An:2016srz}.  The energy spectra of the
deexcitation gammas of gadolinium-155 and 157 are shown in Figure~\ref{fig:gamma155} and Figure~\ref{fig:gamma157}, respectively, for these models.
The gamma model-3 has the hardest gamma spectra.

The 20 available combinations provided by the five neutron scattering model combinations (a-e) and the four gamma models (1-4) are used to estimate $\varepsilon_n$. Model a-1 was used in the previous analyses~\cite{An:2016ses, An:2016srz}.

\begin{table*}[]
\centering
\caption{Summary of five neutron scattering model combinations and four gamma models, including the
efficiency for detecting inverse beta decay neutrons and the $\chi^2$ with 59 calibration source-location points. See text for details.}
\label{tab:models}
\begin{tabular}{lllll}
  \hline\hline
  $\varepsilon_n$,$\chi^2$   & 1. Geant4 native   & 2. Geant4 Phot. Eva.   & 3. Nuclear Data Sheets  & 4. Caltech  \\
  \hline
  a. water, free gas      & 82.23\%, 76.0   & 82.35\%, 	86.4 & 80.56\%,	316 &	82.55\%, 156    \\
  b. water, poly          & 81.75\%, 52.1   & 81.93\%, 	85.1 & 80.42\%,	350 &	82.43\%, 119    \\
  c. poly, poly           & 81.61\%, 56.6   & 82.00\%, 	63.9 & 79.96\%,	389 &	82.00\%, 96.9   \\
  d. poly, free gas       & 82.01\%, 57.7   & 82.28\%, 	79.9 & 80.28\%,	371 &	82.36\%, 115    \\
  e. free gas, free gas   & 84.76\%, 1183   & 84.65\%, 	1273 & 82.70\%,	576 &	85.37\%, 1569   \\
  \hline\hline
\end{tabular}
\end{table*}

\subsection{Data and simulation comparison}
\begin{figure}[]
    \centering
    \subfigure[AmBe excited state at $r$=1.35~m and $z$=-1.35~m]
    {
      \label{fig:CalibSpec1}
      \includegraphics[width=0.45\textwidth]{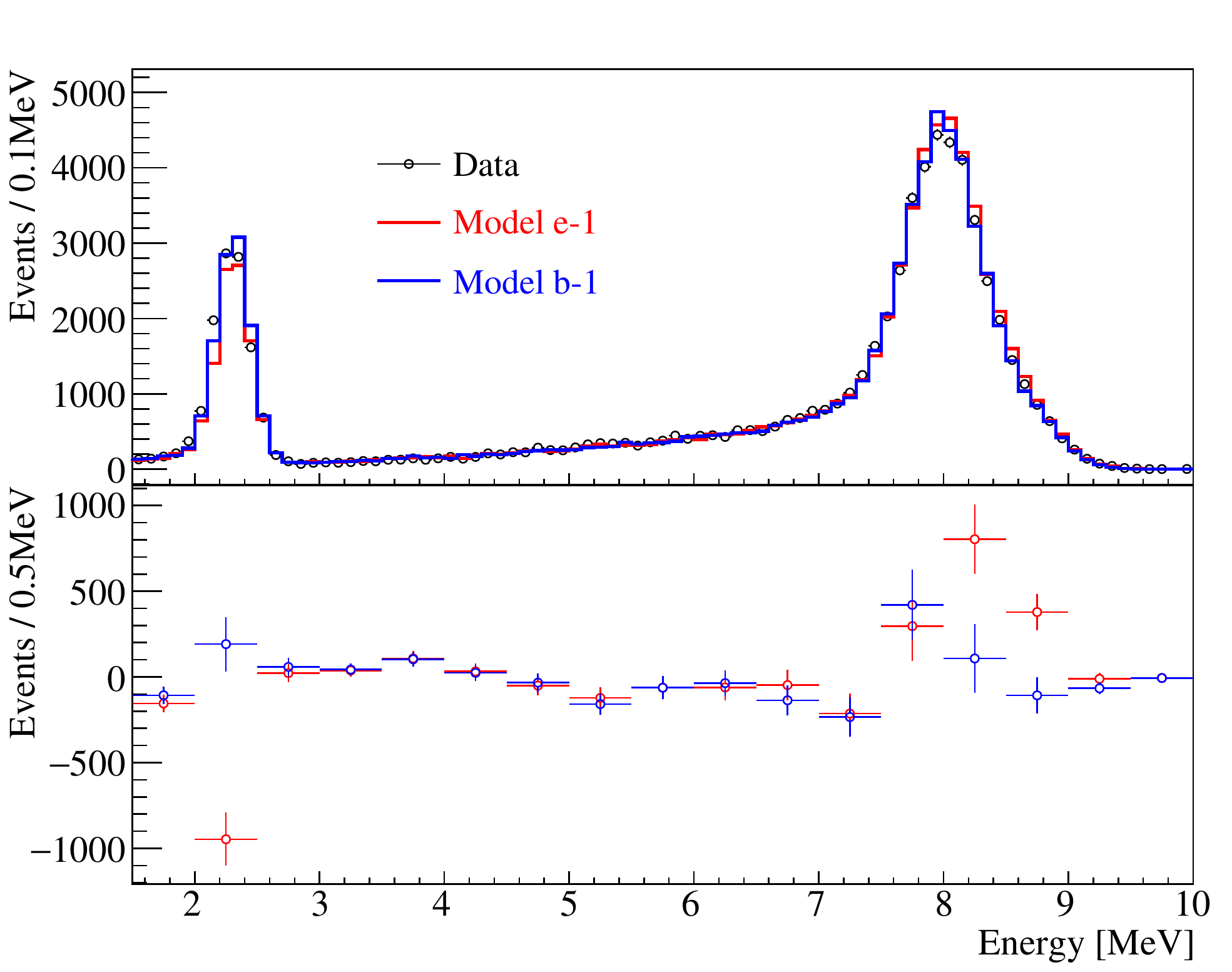}
    }
    \subfigure[AmC ground state at $r$=1.77~m and $z$=0~m]
    {
      \label{fig:CalibSpec}
      \includegraphics[width=0.45\textwidth]{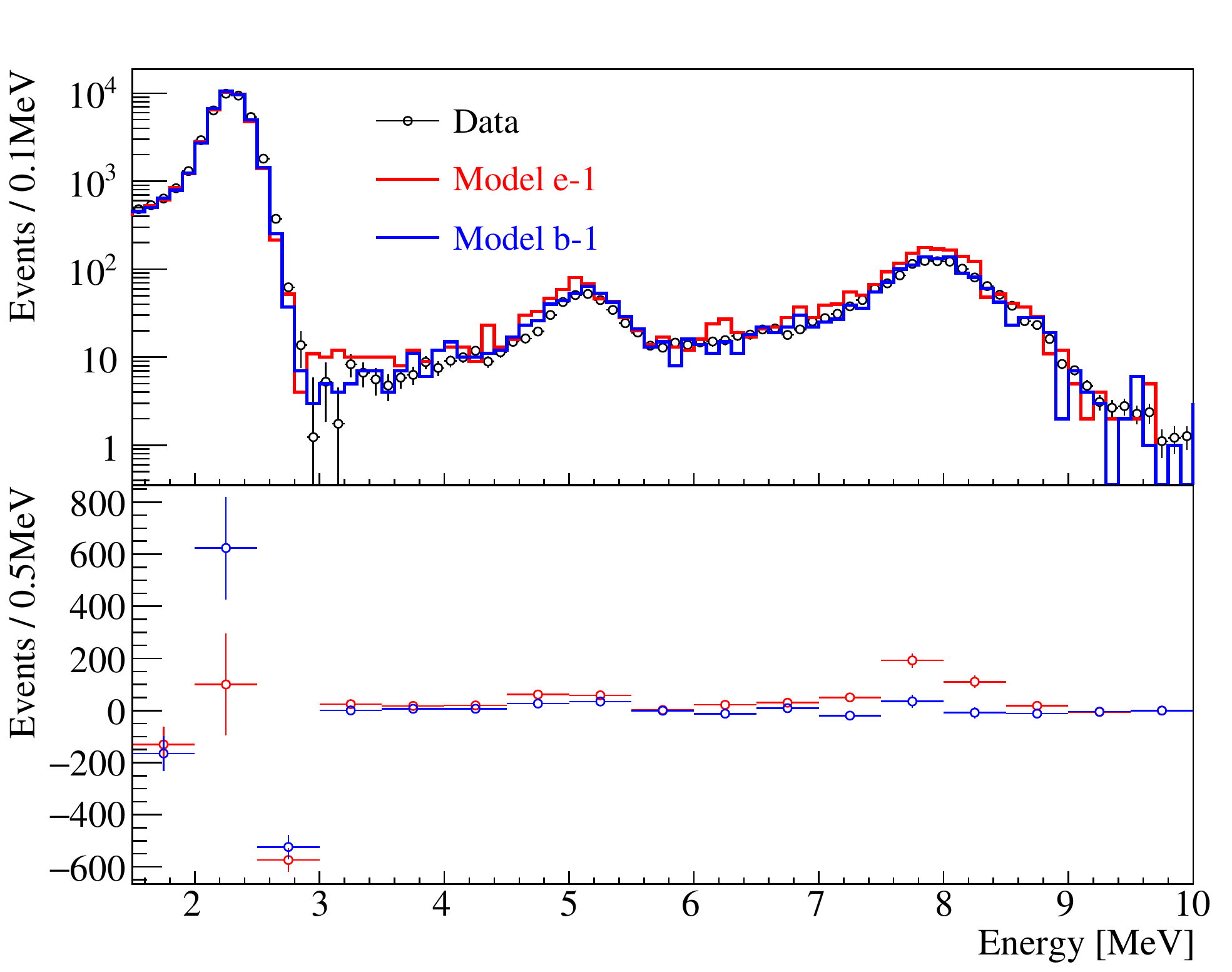}
    }
    \caption{Delayed energy spectra of neutron sources at two
      calibration locations for data and simulation and
      the corresponding residual plots (MC-data).
      Two MC models, b-1 (best fit) and e-1 (rejected), are overlaid.
      Normalization is determined using the integral between 1.5 and 12 MeV.
      The difference of relative gadolinium/hydrogen capture ratio between sub-figure (a) and
      (b) is due to the relative position to the GdLS volume.
      In (b), a weak signal of neutron capture on carbon can be seen. In (b), a mismatch of the energy scales of data and MC at nH peak
      is observed, but our selection cut efficiency is not sensitive to the difference.}
    \label{fig:spectrum}
\end{figure}

For each calibration SLP, a ratio $F$ is calculated.
\begin{equation}
\label{eq:F_ratio}
F=\frac{N([6,12]~\text{MeV})}{N([1.5,12]~\text{MeV})}\ ,
\end{equation}
where $N([6,12]~\text{MeV})$ and $N([1.5,12]~\text{MeV})$ are the numbers of events
with reconstructed delayed energy in the range of $[6,12]~\text{MeV}$ and $[1.5,12]~\text{MeV}$, respectively.
1.5~MeV is chosen to include the hydrogen capture peak.
$F$ is very sensitive to the relative
strength of H vs.~Gd capture peaks and
the containment of the 8~MeV of gamma energy from nGd capture,
and therefore provides a crucial benchmark for the neutron and gamma simulation models.
For the 59 calibration SLPs, a $\chi^2$ is constructed to measure the
overall difference between data and MC predictions,
\begin{equation}
\label{eq:chi2}
\chi^2=(F_{\rm{data}}-F_{\rm{MC}})^T\cdot V^{-1}\cdot(F_{\rm{data}}-F_{\rm{MC}}),
\end{equation}
where $(F_{\rm{data}}-F_{\rm{MC}})$ is a vector with 59 elements of
the difference of $F$ between the data and MC, and $V$ is the
covariance matrix.
For most of the calibration points, the statistical uncertainty is dominant, but for the points near a GdLS boundary (ACU-A top, ACU-A bottom, ACU-B top, or ACU-B bottom), the distance to the acrylic and LS volume is comparable to the neutron drift distance, so they share a large common uncertainty due to the source location $z$ uncertainty.
The $\chi^2$ and $\varepsilon_n$ values for all models are shown in Table~\ref{tab:models}.
The eight combinations with either neutron model-e or gamma model-3 are discrepant
($\chi^2>300$), and therefore are excluded.
The remaining twelve
(4$\times$3) models agree with the data reasonably well with $\chi^2$
in the range of 52.1--156 (``reasonable models'').
The best model with minimum $\chi^2$ is model b-1.
In Figure~\ref{fig:spectrum}, the
delayed energy spectra at two boundary calibration locations from data
are compared to models b-1 and e-1, where model b-1 shows a better agreement with data for $F$.

\begin{figure}[]
\centering
\includegraphics[width=0.5\textwidth]{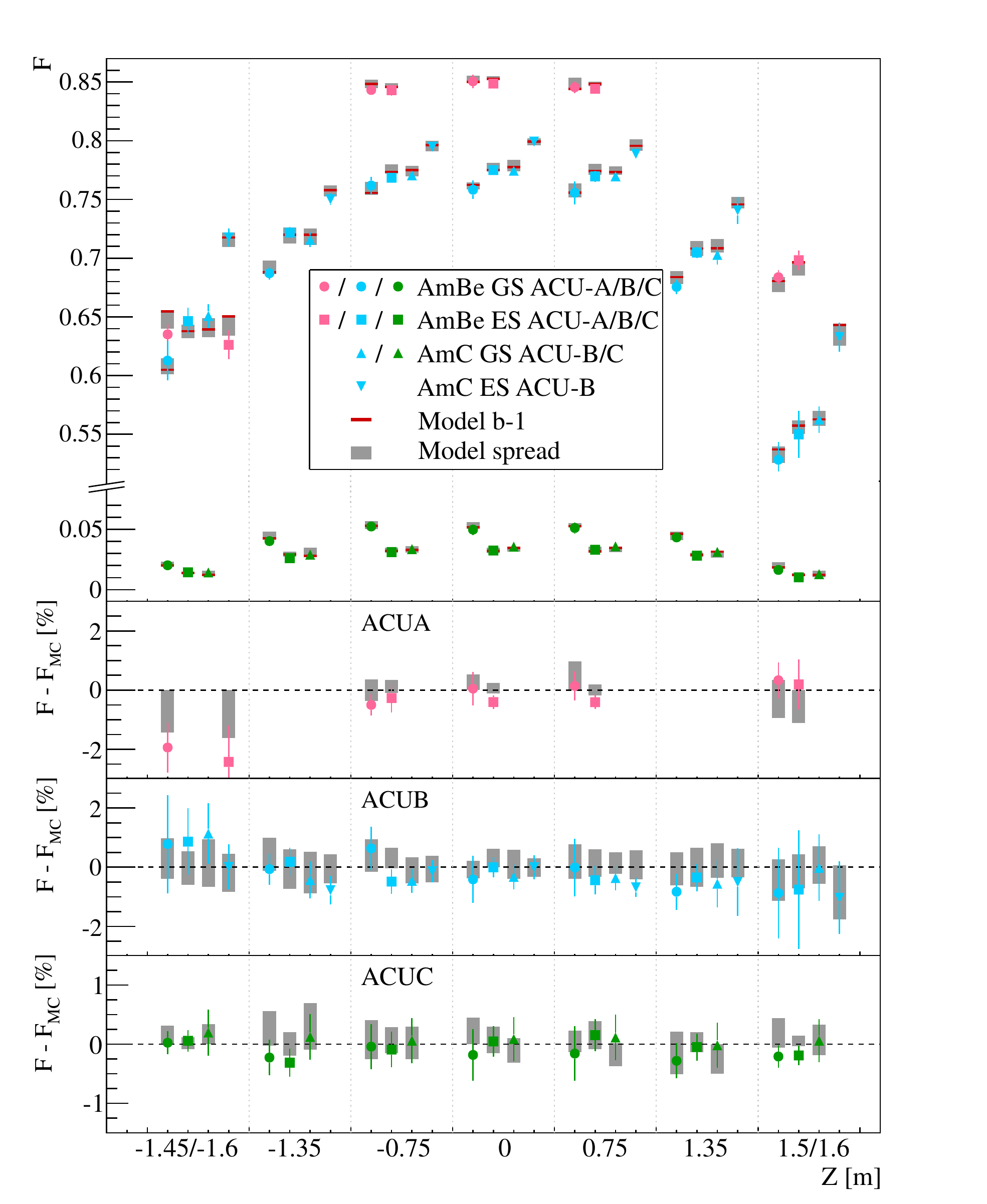}
\caption{ Upper panel: The measured and best MC model (b-1) values of $F$
  for AmC and AmBe (ground and excited states)
  neutrons at three calibration axes and different $z$ positions.
  The model spread of $F$, maximum minus minimum, for each source-location point are also shown.
  The results at $z$=-1.45~m and -1.6~m are plotted together as are 1.5 m and 1.6 m.
  Lower three panels:
  The difference of data and best MC in $F$ along the three vertical
  calibration axes, with the data and MC uncertainties combined. The
  gray bars indicate the spread of the twelve reasonable MC models
  relative to the best model.}
\label{fig:F_comp}
\end{figure}

The data and best MC $F$ values and their differences are shown explicitly in
Figure~\ref{fig:F_comp} for all sources and locations.  The systematic
variations among the twelve reasonable models are overlaid, where the
full spread among them, maximum minus minimum, are plotted as the gray bars.
The variation in $F$ from 1 to 85\% for the 59 data points is due to
the differences in the local geometry and neutron kinetic energy and
 is well reproduced by simulation.
For most points, the best MC model, b-1, reaches an
agreement with data at the sub-percent level, and the residual difference is mostly
smaller than the model spread.
Another quantity,
$F^{\prime}=N([3,4.5]~\text{MeV})/N([3,12]~\text{MeV})$,
is also constructed, which is also sensitive to the gamma model and energy leakage.
The same data-model comparison procedure confirms that
gamma model 3 should be rejected and that gamma model 1 is reasonable.

To further investigate potential effects due to the discreteness of
the calibration sources and the energy difference between neutron
sources and IBD neutrons, we exploited a large sample of IBDs from
data as a special SLP to be compared to the model prediction.
Due to the resolution in position reconstruction, selection
of pure GdLS IBDs is impossible. Instead, a GdLS+LS IBD sample from
all four near site ADs were selected using cuts identical to those used on the
neutron calibration data (Sec. III-A), except that the prompt energy
cut was adjusted to be greater than 3.5~MeV to suppress accidental
background. About 2 million GdLS+LS IBD events in total were
selected.
The measured ratio $F$ is consistent AD to AD with an average of 47.1\%$\pm$0.1\%.
The ratio from model b-1 is 47.0\%, and the full model spread is from 46.7\% to 47.5\%.

\subsection{Neutron detection efficiency determination}
\begin{figure}[] \centering
  \subfigure[AmBe excited state at $r$=1.35~m and $z$=0~m]
  {
    \label{fig:fit1}
    \includegraphics[width=0.45\textwidth]{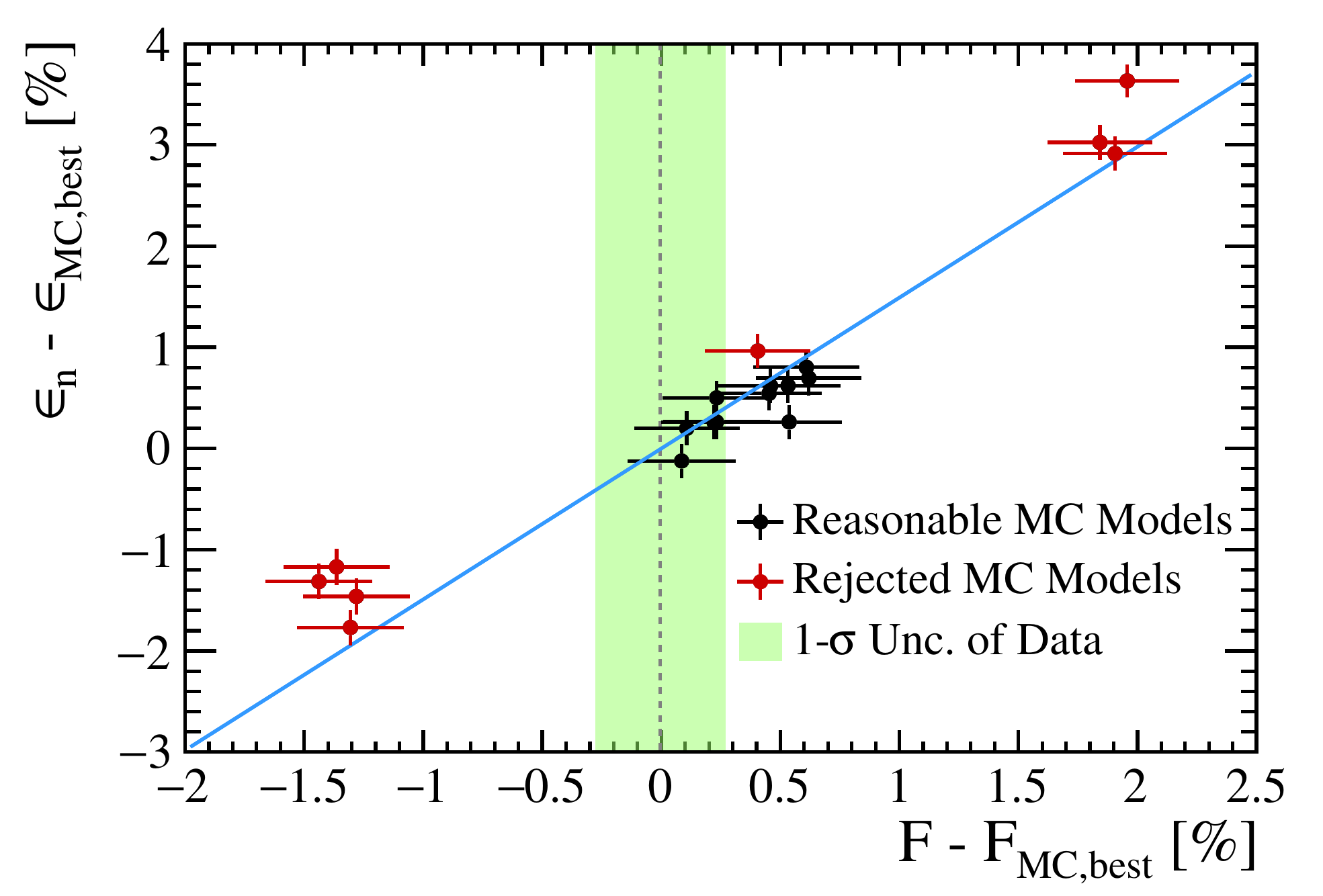}
  }
  \subfigure[AmBe excited state at $r$=0~m and $z$=-0.75~m]
  {
    \label{fig:fit2}
    \includegraphics[width=0.45\textwidth]{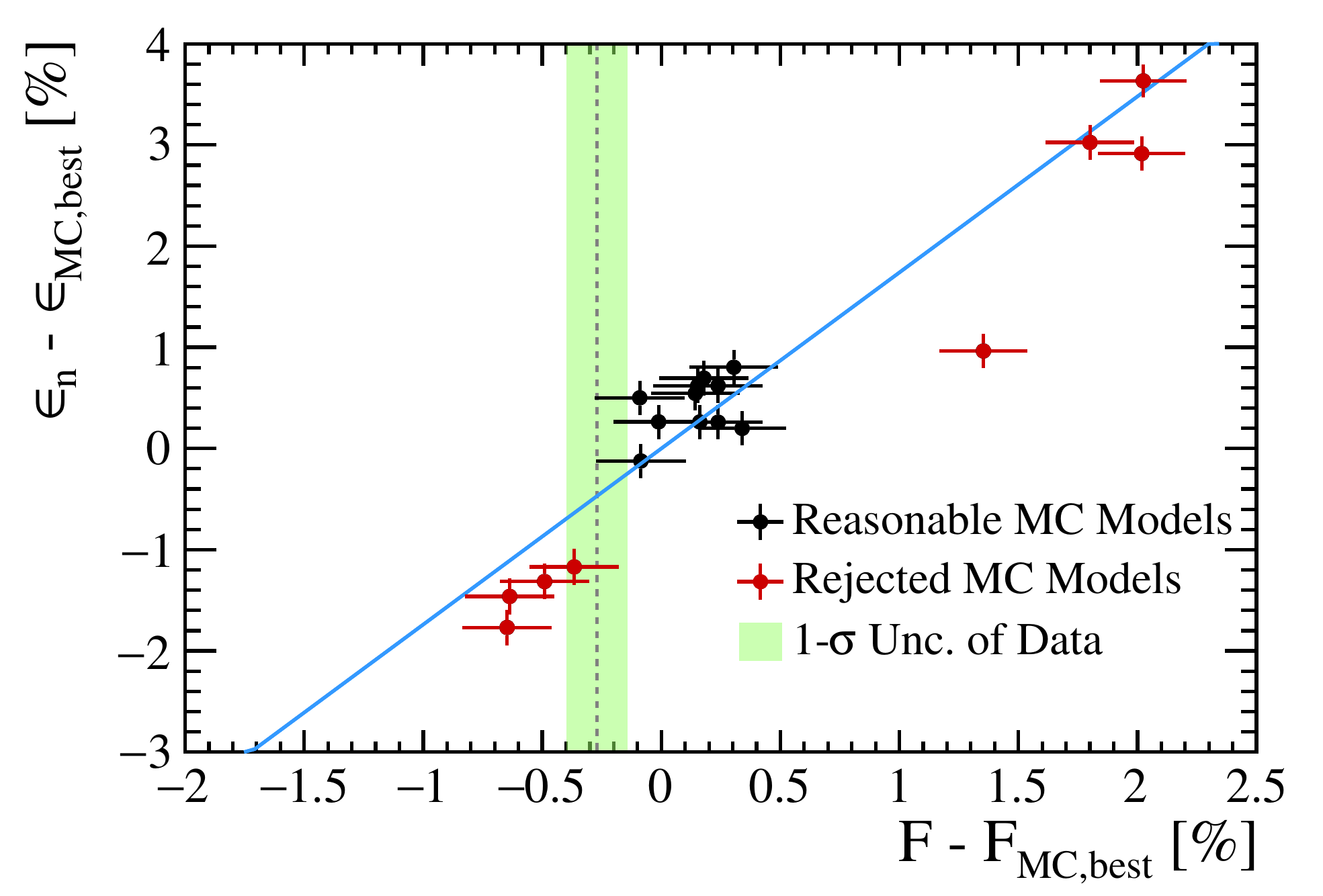}
  }
  \caption{$F-F_{\rm{MC,best}}$  vs.~$\varepsilon_n-\varepsilon_{\rm{MC,best}}$ for the
    20 MC models at two example SLPs.  The vertical dashed line and the shaded band indicates
    the mean and uncertainty of measured $F$ with data.
    The vertical intercept of the linear fit passing through the best MC model (solid blue line)
    with the vertical dashed line gives the corrected value of $\varepsilon_n$.}
  \label{fig:lineFit}
\end{figure}

Each model can give a prediction on $\varepsilon_n$ for IBDs (see Table~\ref{tab:models}).
For the twelve reasonable models, $\varepsilon_n$ ranges from 81.61\%
(model c-1) to 82.55\% (model a-4), and that from model b-1
is 81.75\%.  Instead of taking the prediction as is, one can
translate the data and MC difference in $F$ to a correction to
$\varepsilon_n$, since the two are intrinsically correlated (linear to
the lowest order) through the neutron and gamma models mentioned
above.  In mathematical form, for the $i$th SLP, we have
\begin{equation}
\varepsilon_{n} = c_i \cdot (F_{\rm{data},i}-F_{\rm{MC,best},i}) + \varepsilon_{\rm{MC,best}},
\label{eq:line_reg}
\end{equation}
where $\varepsilon_{\rm{MC,best}}$ is the neutron detection efficiency
given by the best MC model. $c_i$ characterizes the linear correlation
between $F_i$ and $\varepsilon_n$, and can be estimated through a linear regression (fit)
using predicted values of $\varepsilon_{n}$ and $F_i$ from all 20 MC models.
This procedure is illustrated in Figure~\ref{fig:lineFit}.
The eight rejected models were also included here by default
as larger variations in $\varepsilon_n$ and $F$ are allowed.
Excluding them from the fit does not change the result (See Figure~\ref{fig:lineFit} for an example).
In addition to $\varepsilon_{n}$
determined from 59 individual SLPs, a multiple regression procedure
was also applied to model the linear relation between $\varepsilon_n$ and
$F$ from a set of SLPs.
Taking all values of corrected $\varepsilon_n$ into account,
a shift of $-0.27\%$ with a standard deviation of 0.47\% was
obtained, relative to that from the best model (81.75\%).
The standard deviation of 0.47\% is consistent with the model spread on $\varepsilon_n$.

Aside from the model uncertainties, other systematic effects (e.g.
gadolinium abundance, source geometry, absolute energy scale, and
material density variations) have been studied in the MC and found to be negligible.

Based on the above discussions, the final IBD neutron efficiency after
the correction is $\varepsilon_n=((81.75-0.27)\pm0.60)\%=(81.48\pm0.60)\%$,
where 0.60\% is conservatively estimated
using the half-spread, 0.47\%, of $\varepsilon_n$ predicted by all reasonable
models (listed in Table~\ref{tab:models}.)
plus the statistical uncertainty from the MC, 0.12\% (absolute uncertainty).

\section{Antineutrino yield and Comparison with Prediction}
\label{sec:result}
Using the new neutron detection efficiency $\varepsilon_n$ and Eq.~\ref{eq:eff_IBD}, the
IBD detection efficiency $\varepsilon_{\rm{IBD}}$ is $(80.25\pm
0.61)\%$. Using the procedure as in Eq.~\ref{eq:yieldcal}, the mean
IBD reaction yield per nuclear fission is
\begin{equation}
\sigma_f = \yield,
\label{eq:final_yield}
\end{equation}
where the major uncertainties (Table~\ref{tab:Effi}) are from the target proton fraction 0.92\%
(relative uncertainty), dominated by the hydrogen-to-carbon ratio due to
instrumental uncertainty in the combustion measurements, and
reactor-related uncertainty 0.90\% (relative uncertainty) due to reactor power and fission fractions.

The ratio of the yield to the prediction of the Huber-Mueller (or ILL-Vogel) reactor model can be calculated.
The effective fission fractions for four
fission isotopes are defined as
\begin{equation}
  \label{eq:EffFissFrac}
  f_{iso} = \sum_{d=1}^4 \sum_{r=1}^6\frac{N_d^PP_{sur}^{rd}N_r^{f,iso}}{L_{rd}^2}\ /\ \sum_{d=1}^4 \sum_{r=1}^6\frac{N_d^PP_{sur}^{rd}N_r^{f}}{L_{rd}^2},
\end{equation}
where $iso$ refers to one of the four major fission isotopes,
i.e.~$^{235}$U, $^{238}$U, $^{239}$Pu, and $^{241}$Pu, $N_r^{f,iso}$ is
the predicted number of fissions contributed by the $iso^{\rm th}$ isotope in
the $r$th reactor core, and other symbols are defined in
Eq.~\ref{eq:yieldcal}.  In the analyzed data, the effective
fission fractions for the four fission isotopes ($^{235}$U, $^{238}$U,
$^{239}$Pu, and $^{241}$Pu) are determined to be (0.564, 0.076, 0.304,
and 0.056), respectively. The predicted IBD yield is the sum due to all
four isotopes, including corrections due to nonequilibrium effects,
\begin{equation}
  \label{eq:YieldPred}
  \sigma_f=\sum_{iso=1}^4 f_{iso}\int (S_{iso}(E_{\nu})+k_{iso}^{\text{NE}}(E_{\nu})) \sigma_{\rm{IBD}}(E_{\nu})  {\rm{d}}E_{\nu},
\end{equation}
in which $S_{iso}(E_{\nu})$ is the predicted antineutrino spectrum for each isotope given by Huber-Mueller or ILL-Vogel model, $\sigma_{\rm{IBD}}(E_{\nu})$ is the IBD cross section,
and $k_{iso}^{\text{NE}}(E_{\nu})$ corrects for the non-equilibrium long-lived
isotopes.
The calculation integrates over neutrino energy $E_{\nu}$ and the non-equilibrium effect contributes $+$0.6\%~\cite{An:2016srz}.
The ratio between
the measured to predicted reactor antineutrino yield $R$ is $\RHuber$ (Huber-Mueller) and $\RILL$
(ILL-Vogel), where the first uncertainty is experimental and the
second is due to the reactor models themselves.
A breakdown of the experimental uncertainties can be seen in Table~\ref{tab:Effi} (see also Ref.~\cite{An:2016srz}).
The uncertainties from power, spent fuel, and non-equilibrium are treated to be uncorrelated among different reactor cores in the oscillation analysis~\cite{An:2016ses}, and those from fission fraction, IBD cross section, and energy/fission are treated to be correlated.
They are conservatively treated as fully correlated in this analysis, and the total reactor-related uncertainty is 0.9\%.
The total experimental uncertainty has been reduced to 1.5\%,
which is a relative 29\% improvement on our previous study.
The new flux measurement is consistent with the ILL-Vogel model,
but differs by 1.8 standard deviations with respect to the Huber-Mueller model,
with the uncertainty now dominated by the theoretical uncertainty.

\begin{figure}[] \centering
  \includegraphics[width=0.45\textwidth]{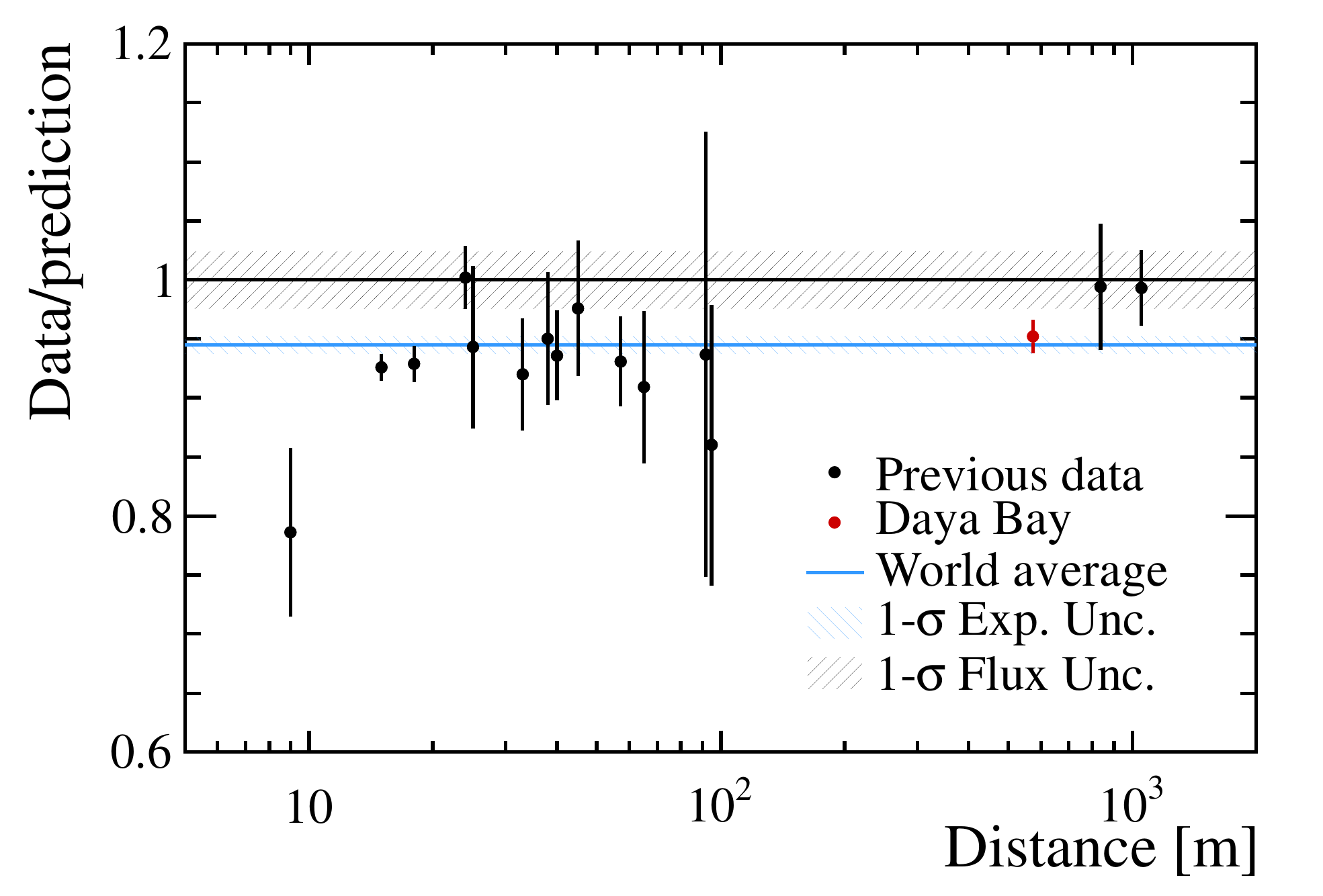}
  \caption{The ratio of measured reactor antineutrino yield to the Huber+Mueller theoretical prediction
  as a function of the distance from the reactor to detector.
  Each ratio is corrected for the effect of neutrino oscillation.
  The blue shaded region represents the global average and its 1$\sigma$ uncertainty.
  The 2.4\% model uncertainty is shown as a band around unity. The measurements at the
  same baseline are combined together for clarity.
  The Daya Bay measurement is shown at the
  flux weighted baseline (578 m) of the two near halls.}
  \label{fig:Ratio-average}
\end{figure}

With the new result, a comparison with the other measurements
is updated using the same method presented in Ref.~\cite{An:2016srz}.
A summary figure is shown in Figure~\ref{fig:Ratio-average}.
The Daya Bay new result on $R$ is consistent with the world data.
The new world average of $R$ is $0.945\pm 0.007\ (\rm exp.) \pm 0.023\ (\rm model)$ with respect to the Huber-Mueller model.
This more precise
measurement further indicates that the origin of RAA is unlikely to be due
to detector effects.

\section{Summary}
In summary, an improved antineutrino flux measurement is reported at
Daya Bay with a 1230-day data set.  The precision of the measured mean
IBD yield is improved by 29\% with a significantly improved neutron
detection efficiency estimation.  The new reactor antineutrino flux is
$\sigma_f = \yield$.
The ratio with respect to predicted reactor antineutrino yield $R$ is
$\RHuber$ (Huber-Mueller) and $\RILL$
(ILL-Vogel), where the first uncertainty is experimental and the
second is due to the reactor models. This yield measurement is consistent
with the world data, and further comfirms the discrepancy
between the world reactor antineutrino flux and the Huber-Mueller
model.

\section*{Acknowledgments}
The Daya Bay Experiment is supported in part by
the Ministry of Science and Technology of China,
the U.S. Department of Energy,
the Chinese Academy of Sciences,
the CAS Center for Excellence in Particle Physics,
the National Natural Science Foundation of China,
the Guangdong provincial government,
the Shenzhen municipal government,
the China General Nuclear Power Group,
the Research Grants Council of the Hong Kong Special Administrative Region of China,
the Ministry of Education in Taiwan,
the U.S. National Science Foundation,
the Ministry of Education, Youth, and Sports of the Czech Republic, the Charles University Research Centre UNCE,
the Joint Institute of Nuclear Research in Dubna, Russia,
the NSFC-RFBR joint research program,
the National Commission of Scientific and Technological Research of Chile,
We acknowledge Yellow River Engineering Consulting Co., Ltd., and China Railway 15th Bureau Group Co., Ltd., for building the underground laboratory.
We are grateful for the ongoing cooperation from the China Guangdong Nuclear Power Group and China Light~\&~Power Company.

\bibliography{refs}

\end{document}